\documentclass{jfm_FR}

\usepackage{graphicx}
\usepackage{newtxtext}
\usepackage{newtxmath}
\usepackage{natbib}
\usepackage{hyperref}
\hypersetup{
    colorlinks = true,
    urlcolor   = blue,
    citecolor  = black,
}

\newcommand{\RomanNumeralCaps}[1]
\linenumbers

\usepackage{amsmath}            
\usepackage{flushend}           
\usepackage{hyperref} 
\usepackage[svgnames]{xcolor}
\usepackage{amssymb}
\usepackage{cellspace} 
\usepackage[figuresright]{rotating}

\usepackage{subfigure}
\usepackage{picture}
\usepackage{color}
\usepackage{mathrsfs}
\usepackage[squaren,Gray]{SIunits}
\usepackage{microtype}
\usepackage{blindtext}
\usepackage{ulem}
\usepackage{makecell}
\usepackage{cancel}
\usepackage{calrsfs}
\usepackage[T1]{fontenc}
\begin{document}

\title{Experimental study of diffusiophoresis in a cellular flow}

\author{Florence Raynal\aff{1}
\corresp{\email{florence.raynal@ec-lyon.fr}},
Menghua Zhao\aff{1}\aff{2}
\and Romain Volk\aff{2}
}
\affiliation{\aff{1}Laboratoire de M\'ecanique des Fluides et d'Acoustique, CNRS UMR 5509, \'Ecole centrale de Lyon, INSA Lyon, Université Claude Bernard Lyon 1, 69134 \'Ecully C\'edex, France
\aff{2}Laboratoire de Physique, ENS de Lyon, CNRS, F-69342 Lyon, France}

\maketitle

\begin{abstract}
In this article, we study experimentally the dispersion of colloids in a two-dimensional, time independent, Rayleigh-B\'enard flow in the presence of salt gradients. 
Due to the additional scalar, the colloids do not follow exactly the Eulerian flow-field, but have a (small) extra-velocity
$\mathbf{v}_{\mathrm{dp}} = D_{\mathrm{dp}} \nabla \log C_s$, where $D_\mathrm{dp}$ is the phoretic constant and $C_s$ the salt concentration. 
Such a configuration is motivated by the theoretical work by \citet{bib:volk_etal2022phoresis}, which predicted enhanced transport or blockage in a stationary cellular flow depending on the value of a blockage coefficient. 
By means of High Dynamical Range Light Induced Fluorescence, we study the evolution of the colloids concentration field at large Péclet number. 
We find a good agreement with the theoretical work although a number of hypotheses are not satisfied,  as the experiment is non homogeneous in space and intrinsically transient. In particular we observe enhanced transport when salt and colloids are injected at both ends of the Rayleigh-B\'enard chamber, and blockage when colloids and salt are injected together and phoretic effects are strong enough. 
\end{abstract}

%\begin{keywords}
%Authors should not enter keywords on the manuscript, as these must be chosen by the author during the online submission process and will then be added during the typesetting process (see \href{https://www.cambridge.org/core/journals/journal-of-fluid-mechanics/information/list-of-keywords}{Keyword PDF} for the full list).  Other classifications will be added at the same time.
%\end{keywords}


%%%%%%%%%%%%%%%%%%%%
\section{Introduction}

The transport of particles or macro-molecules in a flow can be affected by concentration gradients of a scalar field present in the flow such as temperature, solute concentration, or nutrient in the case of living organisms. 
Such phoretic particles do not follow exactly the Eulerian flow-field but have a small drift velocity, the precise expression of which depends on the physical phenomenon involved.  
In the case of charged colloids in water, this extra-displacement is related to  electrokinetic effects, and the phenomenon is called diffusiophoresis: 
due to salt gradients, the drift velocity takes the form $\mathbf{v}_{\mathrm{dp}} = D_{\mathrm{dp}} \nabla \log C_s$, where $D_\mathrm{dp}$ is the diffusiophoretic mobility and $C_s$ is the salt concentration \citep{bib:Derjaguin_etal1961,bib:prieve1984,bib:Anderson1989,bib:ault2024}. 

Although the drift velocity is usually very small as compared to any macroscopic fluid velocity, it has a strong influence on how the phoretic particles spread inside the flow, even at the macro-scale \citep{bib:Maugeretal2016}. 
In particular, it was found that mixing could be delayed or enhanced depending on if the colloids are introduced together with the salt or introduced in salted water:
for instance, \cite{bib:Abecassisetal2009} showed in a $\Psi-$channel that the diffusiophoretic effect could be modelled by a diffusive behaviour, with a positive or negative effective diffusion coefficient much greater than that of the large colloidal particle; in chaotic advection, mixing of colloids was measured using an effective P\'eclet number \citep{bib:Deseigneetal2014,bib:Maugeretal2016,bib:raynal_volk2019}. 
However, the complex behavior of colloidal particles cannot be summed up by a simple diffusive effect; indeed, its origin lies in the compressibility of the drift velocity $\nabla \cdot \mathbf{v}_{\mathrm{dp}}$, which can be set to be positive or negative depending on the configuration \citep{bib:Volketal2014,bib:raynaletal2018,renaud_vanneste_2020,bib:chu_etal2020,bib:chu_etal2021,bib:chu_etal2022}.

Another important situation, often encountered in practical applications and distinct from the previous open flows above, is the case of flows with closed streamlines. 
Indeed most flows have recirculation loops from which a perfect tracer cannot escape if the flow remains steady. 
In particular, the transport of a passive scalar in an array of vortices with closed streamlines was the focus of several theoretical and experimental studies 
\citep{bib:shraiman1987,soward_1987,young1989anomalous,solomon1988passive}. 

In such a configuration, it has been shown that advection-diffusion of Brownian particles leads to diffusive dynamics at long times with a ratio of their effective diffusivity to the molecular diffusivity growing as the square root of the P\'eclet number, $\mathit{Pe}=UL/D$, where $U$ is the velocity scale, $L$ the size of the recirculation cell, and $D$ the molecular diffusivity of the particles. 
More recently, this study has been extended to the theoretical case of joint transport of salt and phoretic particles \citep{bib:volk_etal2022phoresis}. 
In this work, salt inhomogeneities are forced by a constant salt gradient $\mathbf{G}$, leading to strong salt variations in between vortices; 
phoretic particles have a linear drift velocity of the form $\mathbf{v}_{\mathrm{drift}} = \alpha \nabla C_s$ and are initially homogeneously released. 
By means of numerical simulations performed in both Eulerian and Lagrangian framework, in the limit of realistic large P\'eclet numbers for both salt and colloids, the authors could address how colloids move or regroup in the flow under the action of advection, diffusion and phoresis.
The study showed the existence of two regimes, both very different from the diffusive-like behaviour observed in previous studies, depending on the value of a parameter $\displaystyle R=\frac{\alpha G L}{2\sqrt{D_cD_s}}$, where $D_s$ and $D_c$ are the diffusion coefficients of salt and colloids respectively: 
\begin{enumerate}
    \item[$(i)$] For $R\ll 1$, the transport of colloids was strongly enhanced, leading to a mean velocity $v_m\sim \alpha G \sqrt{\mathit{Pe}_s}$, where $\mathit{Pe}_s=UL/D_s$ is the salt P\'eclet number; this velocity is the much larger than the phoretic velocity $\alpha G$ that the particles would have in the absence of flow. This regime is called ``enhanced transport'' in the following;
    \item[$(ii)$] For $R\gg1$, when phoretic effects are very strong compared to diffusion, instead of moving even more rapidly in the direction of the mean salt gradient, the colloids are trapped in their cells.  
    This regime is referred to as ``blockage'' in the following, and $R$ is called ``blockage coefficient''. 
\end{enumerate}

However, these theoretical findings were obtained in the stationary regime with periodic boundary conditions and a  mean salt gradient constant in space, so that phoretic effects could build up in time without any constraint. One may then wonder if these two regimes can be observed experimentally after injection of salt and colloids in some bounded convective flow, a problem of dispersion that is intrinsically transient and non homogeneous in space.

In this article, we propose to prove experimentally the existence of the two regimes of enhanced transport and blockage. 
To this end, we use the same phoretic particles as in \cite{bib:Maugeretal2016}, \textit{i.e.} charged colloids which are attracted toward regions of high salt concentration.
This makes it impossible to create the cellular flow by electro-convection \citep{bib:bergougnoux_etal_2014} as it would require already salted water, which could screen the diffusiophoretic effects. 
Rather, we use the experimental configuration of \citet{solomon1988passive}, which takes place in a long horizontal Rayleigh-Bénard cell, where the flow is time-independent and composed of an alignment of steady convection rolls in the closed  chamber, see figure \ref{fig:cell_setup}. 
In order to study how the injected colloids or salt spread  inside the chamber, we follow their instantaneous concentration field using High Dynamical Range Light Induced Fluorescence (HDR LIF). 
We compare our experimental findings with the theory, and recover the two regimes, in particular the counter-intuitive regime of blockage.

The article is organized as follows: in Section \ref{section:setup}, we describe the experimental setup and the flow inside the cell; we also detail the multi-exposure times technique used to build the HDR images obtained by LIF, and explain how this was corrected to account for the attenuation of laser light when converting recorded images into concentration fields. 
In Section \ref{section:single} we characterize the transport of a single species in the flow.  
 We start with fluorescein, whose diffusion coefficient is close to that of salt, and extend these results to salt, whose concentration field is required to apply our theoretical results. Then we study the case of transport of colloids in the absence of salt, therefore without phoretic effects, in order to get a reference for comparison with cases where salt is present.
Section \ref{section:diffusio}  is devoted to the phoretic cases, when both colloids and salt are present in the flow. 
We consider two configurations, one where colloids and salt are injected on the two sides of the chamber (salt-out configuration), the second where colloids and salt are injected together on one side of the chamber (salt-in configuration). 
We first summarise the theoretical results obtained in \cite{bib:volk_etal2022phoresis},in order to transpose them to the experiment afterwards. 
We then discuss how the regimes of enhanced transport and blockage are obtained. Finally, in the last section, we conclude and summarise the work. 

%%%%%%%%%%%%%%%%%%%%%
\section{Experimental set-up}
\label{section:setup}

\begin{figure}
      \includegraphics{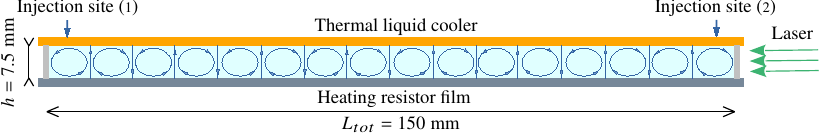}
    \caption{
    The flow in the Rayleigh-B\'enard cell (with length $L_\mathit{tot}=\unit{150}\milli\meter$, height $h=\unit{7.5}\milli\meter$, width $w=\unit{15}\milli\meter$) is set above onset of the first instability and below second instability using a heating thermal Kapton heater
    at the bottom and a liquid cooler at the top. Two injection sites are located on both sides of the cell. 
    The vertical walls are made of transparent acrylic glass. 
    The fluorescent species (whether colloids or fluorescein) is injected through site (1), and its dispersion is followed using a laser sheet.
    }
    \label{fig:cell_setup}
\end{figure}

Mixing takes place in an elongated Rayleigh-Bénard (RB) cell whose geometry is identical to that of \cite{solomon1988passive}. 
Its horizontal dimensions are $L_{tot}=15\,\centi\meter$ and $w=1.5\,\centi\meter$, and its height is $h=0.75\,\centi\meter$, as shown in figure \ref{fig:cell_setup}. 
The vertical walls of the chamber are made of transparent acrylic glass in order to perform optical measurements, while top and bottom covers are composed of brass to ensure a good thermal contact with the fluid.

For concentration measurements, we operate with fluorescent species: fluorescein, whose diffusion coefficient is close to that of the salt used for diffusiophoretic effects (LiCl), and fluorescent colloids, carboxilate microspheres (Invitrogen) of diameter $0.2\,\micro\meter$ already used in \cite{bib:Maugeretal2016}.  
The colloids are isodense particles, with negligible inertia.
The fluorescent species is introduced through injection hole ({\scriptsize 1}), opposite the laser. 

The local concentration of a generic fluorescent species is hereinafter referred to as $C(x,y,t)$.
Because we are mostly interested in the spreading in the $x-$direction, for simplicity we extend the notation $C$ for $y-$averaged concentration profiles, \textit{i.e.} $C(x,t)=1/h\int_{y=0}^{y=h} C(x,y,t)\,dy$.

\subsection{Flow characterization}

In order to generate a convective flow, we create a temperature difference between the top and bottom: a fixed temperature of $T=26\degreecelsius$ is imposed on the top cover with a water circulation and a thermal bath, and we set the heat flux through the chamber using a thin $7.5$ W Kapton heater placed under the bottom cover.  
The heating resistive sheet, which is connected to a DC supply, is compressed between the bottom part and (an insulating) PVC block to ensure that most of the heat flux is directed toward the liquid.
In order to tune the temperature difference, and more especially set the bottom temperature, top and bottom temperatures are monitored using a picolog acquisition card connected to 4 k-type thermocouples inserted in the top and bottom part, in close contact with the fluid.  
Such thermal driving proved to ensure fine tuning of the temperature difference, $\Delta T \in [0.9, ~1.5]\degreecelsius$, between top and bottom part with good stability. Finally, the whole setup is placed in a temperature regulated room to increase the thermal stability of the setup, which is required since the measurements of mixing last about 10 hours.

For all experiments, we use pre-boiled distilled water to either fill the chamber or prepare the salt/colloidal solutions so as to prevent the formation of air bubbles during the long heating of the RB chamber. 
When operating, the chamber is filled with liquid using two holes drilled on the top cover connected to syringe pumps. 
When injecting dye or colloids, the injection rate is set to be less than $20\,\micro\liter\per\minute$ to avoid any strong perturbation of the convection flow.
During an injection process the lid of the opposite injection site is kept open so as to allow excess water to escape.

\begin{figure}
\!\!\!\!\!\!\!\!\!\!\!\includegraphics{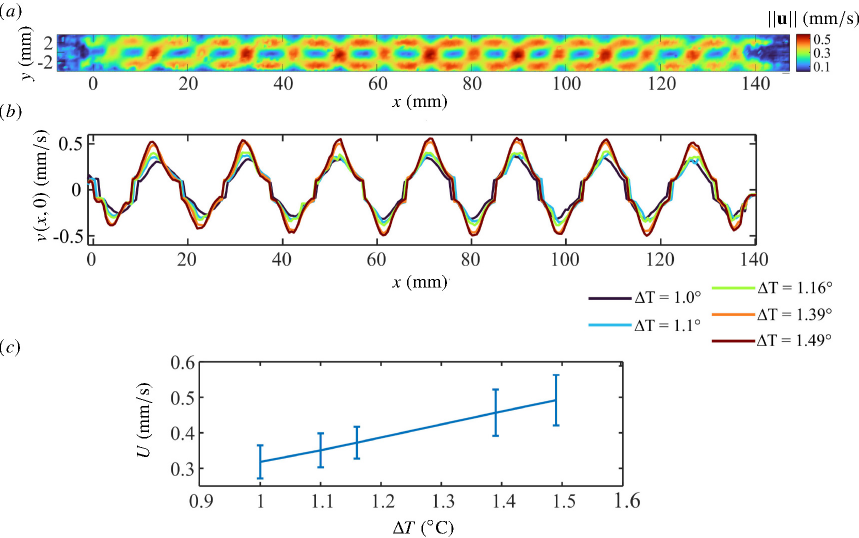}
 \caption{{$(a)$ Magnitude $||\mathbf{u}||$ of the velocity field measured by PIV for $\Delta T = 1.37$ $^\circ$C. $(b)$ Vertical velocity profile along the center line inside the RB chamber for several values of the temperature difference $\Delta T$, from $1.00\,\degreecelsius$ to $1.49\,\degreecelsius$. $(c)$ Maximum of the velocity magnitude $U$ as a function of $\Delta$T.} 
 }
\label{fig:figure2}
\end{figure}

For such thermal forcing, the flow regime is controlled by the Rayleigh number $\displaystyle \mathrm{Ra}=\frac{\gamma g \Delta T h^3}{\kappa \nu}$, where $\gamma$ is the thermal expansion of water, $g$ the gravity, $\kappa$ the thermal diffusivity and $\nu$ the kinematic viscosity of the fluid. 
The flow is found to be convective when $\mathrm{Ra}$ is larger than the critical value $\mathrm{Ra}_\mathit{cr}=1708$. 
In the present case where $h=0.75\,\centi\meter$ and the fluid is water, this gives $\mathrm{Ra}/\mathrm{Ra}_\mathit{cr} \simeq C \Delta T$ where $C=9.2\,\kelvin^{-1}$. 
Given the range of temperature difference, the set-up is expected to produce steady laminar flows as the threshold for the first bifurcation toward a time-dependent flow corresponds to $\mathrm{Ra}/\mathrm{Ra}_c  \approx18$ \citep{solomongollub2}.

This was verified by performing Particle Image Velocimetry (PIV) measurements of the flow field for increasing values of the temperature difference $\Delta T$. 
To achieve this we use a CW laser (Coherent Genesis MX SLM-Series, $1\watt@488\,\nano\meter$),  a cylindrical lens and a $250\,\micro\meter$ slit to form a thin vertical laser sheet in the middle of the RB chamber (see figure \ref{fig:cell_setup}); the flow is seeded with $10$ $\mu$m silver-coated hollow glass spheres from Dantec. 
Images are then recorded at 10 fps using a XIMEA camera and we perform the correlation of pairs of images using a home-made software written in Matlab. 
The result for the velocity magnitude, averaged in time over $1$ minute, is shown in figure \ref{fig:figure2}\,$a$ where the typical signature of well organized convection rolls is visible. 
As shown in Figure \ref{fig:figure2}\,$b$, which displays a horizontal profile of the vertical velocity at mid-depth of the cell, the spatial organization of the flow remains the same for all values of $\Delta T$ in the range $[1,1.5]\degreecelsius$, and consists of about 16 counter-rotating rolls with width $L \approx 9.4\,\milli\meter$, whose positions are pinched within the chamber. 

In the case of Rayleigh-B\'enard flow, the bifurcation is of the Hopf type, meaning that the maximum vertical velocity, denoted by $U$, is such that $U \propto \sqrt{\Delta T-\Delta T_\mathit{cr}}$ close to the threshold, where $\Delta T_\mathit{cr}$ is the top-bottom temperature difference at the threshold. 
Here we operate quite far from the threshold, with a limited range of accessible values of the Rayleigh number, so that the flow amplitude rather tends to increase linearly with the temperature difference (figure \ref{fig:figure2}\,$c$). 
By fitting linearly the averaged value of $U$ as a function of $\Delta T$, we find $U=0.36\times \Delta T-0.04\,\milli\meter\per\second$ so that the Reynolds number of the flow based on the chamber height, $Re=Uh/\nu$ where $\nu=10^{-6}\,\meter^2\per\second$ is the kinematic viscosity of water, is such that $2.5\le Re\le 4.5$ for all experiments.

\subsection{Multi time exposure concentration field measurements}

When a chemical species is injected on one side of the chamber, it is advected and diffused through the convection rolls. 
Such mixing process in a steady and well organized flow is rather inefficient so that the concentration field spans a high dynamical range during the whole run, with large concentration differences and strong concentration gradients located near separatrixes \citep{solomon1988passive}. 
It is therefore very hard to perform direct optical measurements of such concentration field, even with a high quality camera. 
For that purpose, we use a variant of the setup described in  \cite{bib:Maugeretal2016} and perform planar Laser-Induced-Fluorescence (PLIF) measurements using a multiple time exposure technique.

\begin{figure}
     \includegraphics{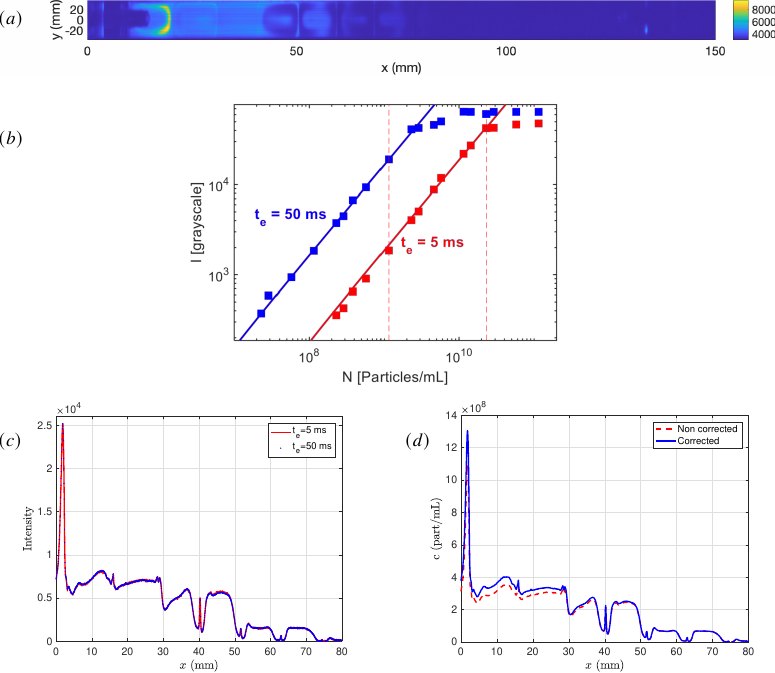}
    \caption{$(a)$ Raw image obtained using the LIF setup an exposure time $t_e=50\,\milli\second$ $10\,\hour$ after colloid injection. 
    $(b)$ Camera calibration curves: ($\blacksquare$) recorded intensity measured with increasing homogeneous fluorescein concentration in $2.5\;\milli\liter$ cuvettes for two exposure times $t_e=5\,\milli\second$ and $50\,\milli\second$. The straight lines correspond to linear fits derived in the linear range $I \leq 44000$ ($60\%$ of the dynamical range). 
    $(c)$ Comparison of $x$-profiles of recorded intensity averaged over the vertical coordinate $y$ and two exposure times. ($\cdot$) raw intensity profile measured with $t_e=50$ ms. ($-$) intensity profile measured with $t_e=5$ ms, transformed using the linear transformation $\mathcal{T}[I]=\gamma I + \kappa$ with $\gamma= 10.2$ and $\kappa = - 951$. $(d)$ $x$-profiles of concentration averaged over the vertical coordinate $y$ built using the HDR procedure and calibration curves obtained in the chamber with known (homogeneous) colloid concentration. $(--)$ without correction for exponential attenuation through the chamber. $(-)$ with compensation for exponential attenuation.}
    \label{fig:calib}
\end{figure}

We use the same optical setup as for the PIV measurements to create the laser sheet, with almost homogeneous light intensity in the chamber. The fluorescence signal is recorded using a $16$-bit camera (Nikon D700, $4200\times2800$ pixels$^2$) equipped with a TAMRON $90\,\milli\meter$ macro lens and a band reject filter (notch $488 \pm 12\,\nano\meter$) corresponding to the laser wavelength. 
For all experiments, the laser power is set to $195\,\milli\watt$ and the ISO sensitivity to its lowest value to avoid noise. 
After recording, images are cropped to fit the aspect ratio of the cell so that their actual size is $3249 \times 181$ pixels$^2$ in $x$ and $y$ directions with a spatial resolution of about $44\,\micro\meter$. 
A typical raw image recorded with an exposure time $t_e=50\,\milli\second$ several hours after colloid injection on the left side injection site  (1) is displayed in figure \ref{fig:calib}\,$(a)$.

\subsubsection{Camera calibration} 

In order to calibrate the setup, one has to relate the recorded intensity on raw images to the local concentration. 
This was done by using an ensemble of solutions  obtained by dilution in $2.5\,\milli\liter$ cuvettes, with known fluorescein or colloid concentration. For each solution, we have recorded images for various exposure times in the range $t_e \leq 500\,\milli\second$ in order to check for the linearity of the measurement. 
The result for the colloids and $t_e=5\,\milli\second$ and $t_e=50\,\milli\second$ is displayed in figure \ref{fig:calib}$b$: the response of the camera is found to be linear when the gray level is smaller than $44000$, which is approximately two thirds of its total dynamical range ($2^{16}$). When operating at low enough intensity, we could verify that the measured intensity is linear in colloid concentration (solid lines in figure \ref{fig:calib}$b$), laser power, and exposure time as expected for a CMOS camera \citep{wang2018linearity}. 

\subsubsection{High Dynamical Range measurement} 
Mixing lasts about $10$ to $20$ hours in the RB chamber so that we use two exposure times, $5\,\milli\second$ and $\tau_2=50\,\milli\second$, and record pairs of images every $100$ seconds; we then build a High Dynamical Range (HDR) image from both measurements. 
To do this we use the linear transformation, $\mathcal{T}[I]=\gamma I+\kappa$, such that $I_{50}(x,y) = \mathcal{T}[I_5(x,y)]$ linking both intensities measured with the two exposure times. Indeed, using such relation one can superimpose horizontal intensity profiles obtained from $I_{50}$ and $\mathcal{T}[I_5]$ in the linear regime as shown in figure \ref{fig:calib}\,$(c)$\footnote{From the data we obtain $\gamma=10.2$, which is very close to the ratio between the two exposure times, and $\kappa=-951$ which depends on the electronic offsets of the camera at $t_e=50\,\milli\second$ and $t_e=5\,\milli\second$, since the exposure times are switched after performing a black reference prior to the experiment.}.

 Potential non-linearities can be eliminated by defining an HDR image as $I_\mathit{HDR}(x,y)=I_{50}(x,y)$ everywhere where $I_{50}$ is smaller than $60\%$ of the sensor dynamical range, and $I_\mathit{HDR}(x,y)=\mathcal{T}[I_{5}(x,y)]$ in any other location. 
This ensures that we have a low level of noise in the measurement while using the whole dynamical range of the camera sensor.

Note that the choice of the threshold and values of $t_e$ have a weak impact on the final result, but the present choices are a good compromise given the laser power and typical maximal colloid concentration in the chamber. 
After combining both raw images taken with two exposure times to build a HDR image, we get rid of ambient illumination by subtracting a background reference image taken with the laser off before starting the experiment. 
We finally transform these HDR images into concentration fields using concentration-intensity calibration curves while taking into account absorption of light in the chamber as explained in the next section.

\subsubsection{Compensation for spatial decay of incoming light} 
Given the low concentrations of fluorescent species $C$ used here, the light intensity scattered per unit volume of fluid in the laser sheet is proportional to the local concentration $I_\mathit{scat}(x,y,t)=\sigma \cdot C(x,y,t) \cdot I_0(x,y,t)$ where $\sigma$ is a constant specific to the colloids and $I_0(x,y,t)$ the incoming light at location $(x,y)$ and time $t$. 
Although the precise values of $\sigma$ and $I_0$ are not needed as we will work with normalized quantities in the following, light scattering may affect the measurement due to the spatial decay of incoming light $I_0(x,y,t)$, corresponding to the Beer–Lambert law \citep{bib:swinehart1962}. In the present case we inject colloids at the location $x=0$ and propagate the laser beam from $x=L_{\mathit{tot}}$ toward $x=0$ so that the local Beer-Lambert law reads
\begin{equation}
\frac{d \log I_0(x,y,t)}{dx} = \sigma\, C(x,y,t)\,. 
\end{equation}
This relation can be integrated from $x$ to $L_{\mathit{tot}}$ to give 
\begin{equation}
I_0(x,y,t) = I_0(L_{tot},y,t) \cdot \exp \left(-\sigma \int_x^{L_{tot}} C(x',y,t) dx' \right) \,,
\end{equation}
where $I_0(L_{tot},y,t)$ is a constant here as the incoming Laser light has an almost homogeneous intensity in the $y$ direction at $x=L_{\mathit{tot}}$, with a constant amplitude in time. 
Due to the spatial decay of the laser intensity along the chamber, the recorded intensity $I_\mathit{HDR}$ is not simply proportional to the concentration. 
However, $I_\mathit{HDR}$ is still proportional to the scattered light intensity if the concentration field is well resolved spatially with a camera operating in the linear regime. 
Assuming this is the case here we then have:
\begin{equation}
I_\mathit{HDR}(x,y,t) = A \cdot \,C (x,y,t) \cdot \exp \left(-\sigma \int_x^{L_{tot}} C(x',y,t) dx' \right)
\label{eq:I_f(A,c,sigma)}
\end{equation}
where $A$ is a multiplicative constant.

To recover an unbiased estimate of the concentration field it is important to accurately measure $\sigma$ and $A$. 
This is achieved by recording images in the case of a homogeneous concentration of  fluorescent species for which a well defined exponential decay was found, leading to $\sigma=1.27 \cdot 10^{-11}\,\milli\meter^2$ and the multiplicative constant $A=2.3\times 10^{-5}\, \mathrm{[graylevel]\cdot part}^{-1}\cdot\milli\liter$ for $t_e=50$ ms. 
Note that $A$ is not specific to the fluorescent species, but changes when the laser power, the exposure time, or the optical setup are modified. 
Finally, because the exponential decay is small here and can be considered as a first-order correction, the concentration field is finally obtained from equation \eqref{eq:I_f(A,c,sigma)} as
\begin{equation}
C(x,y,t) = \frac{I_\mathit{HDR}(x,y,t)}{A} \cdot \exp \left(\frac{\sigma}{A} \int_x^{L_{tot}} I_\mathit{HDR}(x',y,t) dx' \right)\,.
\end{equation}
Using this procedure, the global mean concentration of the fluorescent species,
\begin{equation}
\langle C\rangle=\frac{1}{L_{tot}\times h} \int_0^{L_{tot}} \int_0^h C(x,y,t) \,dx\,dy=\frac{1}{L_{tot}} \int_0^{L_{tot}} C(x,t) \,dx
\end{equation}
is checked to be a conserved quantity. 
The HDR profile of figure \ref{fig:calib}\,$c$  is transformed into a concentration profile as plotted in \ref{fig:calib}\,$d$ where we compare the raw profile without exponential correction and the corrected profile. 
The correction is cumulative from right to left and is about $25\%$ close to the injection point. 
Although such correction may appear significant, shining the laser from the opposite side (Fig. \ref{fig:cell_setup}) reduces its impact to values smaller than $5\%$ on first and second moments of the concentration field
\begin{equation}
\langle x^p \rangle = \frac{1}{L_\mathit{tot}}\,
\int_0^{L_\mathit{tot}} x^p \,\frac{C(x,t)}{\langle C\rangle}\, dx\,, \qquad p=1,2
\label{eq:first_second_moments_generic}
\end{equation}
which are the quantities we will consider in the sequel.

In the following, we apply all the methods described above to follow species in the cellular flow, first without diffusiophoresis, then when diffusiophoresis comes into play.

%%%%%%%%%%%%%%%%%%%%%%%%%%%%%%%%%%%%%%%%%%%%%%%%%%%%%
\section{Single scalar transport in the cellular flow} \label{section:single}
We first characterize the motion of a single species in the cellular flow. 
In practice, the LiCl salt chosen is not fluorescent, so as to avoid any problem when dealing with diffusiophoresis (which involves both colloids and salt). 
In order to evaluate the salt profile, we begin with fluorescein, a species we have already used \citep{bib:Maugeretal2016} since its diffusion coefficient is very close to that of salt, see table \ref{Tab:1}, and then we transpose to LiCl.

Quantities for fluorescein, salt and colloids are labeled with an index $f$, $s$ and $c$ respectively.
For example, we shall denote the concentrations of fluorescein, salt and colloids as $C_\mathit{f}$, $C_\mathit{s}$ and $C_c$, and their diffusion coefficients as $D_\mathit{f}$, $D_s$, $D_c$, and $\mathit{Pe}_\mathit{f}=UL/D_\mathit{f}$, $\mathit{Pe}_s=UL/D_s$, and $\mathit{Pe}_c=UL/D_c$ their corresponding P\'eclet numbers.
Because the length $L$ of the cells is fixed in all experiments, these P\'eclet numbers are directly proportional to velocity, hence increase with the temperature difference $\Delta T$ between the bottom and top. Therefore for simplicity a given experiment is characterized by $\Delta T$ afterwards.
Relevant experimental parameters are summarized in table I, where the diffusion coefficient of colloids is derived from Stokes–Einstein equation and the P\'eclet number $Pe=Uh/D$ based on the chamber height $h$ is given for extreme temperature differences, $\Delta T = 1.0\degreecelsius$ and $\Delta T = 1.5\degreecelsius$.
\begin{table}
\begin{center}
\begin{tabular}{|l|c|c|c|}
\Xhline{0.5pt}
 & $D$ [$\meter^2\per\second$] & $\mathit{Pe}$ at $\Delta T = 1.0\degreecelsius$ & $\mathit{Pe}$ at $\Delta T = 1.5\degreecelsius$\\
\Xhline{0.5pt}
$\text{ Fluorescein }$ & $ 0.42\times 10^{-9}$ & $5.63\times 10^{3}$ & $8.81\times10^3$\\
$\text{ Salt (LiCl)} $ &  $1.3\times 10^{-9}$ & $1.84\times 10^{3}$ & $2.88\times10^3$\\
$\text{ Colloids }$ &  $2.2\times 10^{-12}$ & $2.26\times 10^{6}$ &  $3.53\times 10^6$ \\
\Xhline{0.5pt}
\end{tabular}
\end{center}
\caption{Diffusion coefficient for fluorescein, salt (LiCl) and colloids, and corresponding P\'eclet number of the flow for two extreme temperature differences $\Delta T = 1.0\degreecelsius$ and $\Delta T=1.5\degreecelsius$. We shall note $D$ a generic diffusion coefficient and $D_\mathit{f}$, $D_s$, $D_c$ the corresponding values for fluorescein, salt, and colloids respectively. We will also denote $Pe=UL/D$ a generic P\'eclet number, and write $\mathit{Pe}_\mathit{f}=UL/D_\mathit{f}$, $\mathit{Pe}_s=UL/D_s$, $\mathit{Pe}_c=UL/D_c$ the P\'eclet numbers for the three species.}
\label{Tab:1}
\end{table}

\subsection{Fluorescein profiles and effective diffusivities}

\begin{figure}
    \includegraphics{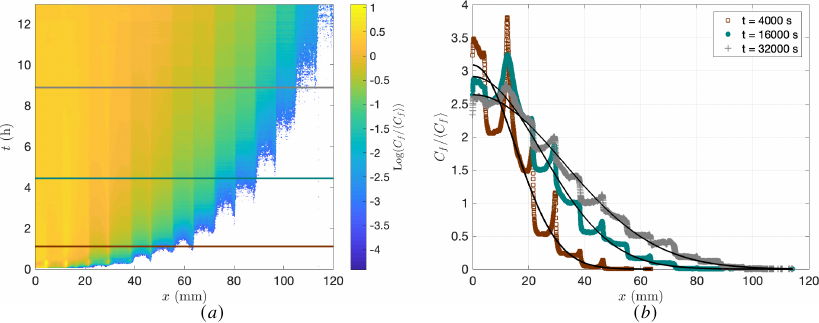}
    \caption{
    $(a)$ Fluorescein concentration profile as a function of time for $\Delta T=1.10\pm0.06\degreecelsius$ ($\mathit{Pe}_\mathit{f}=6\,270$); $(b)$: three concentration profiles extracted from figure \ref{fig:fluo_c_x_t0}\,$a$ (corresponding to the three horizontal lines on this figure) for $t=\unit{4000}\sec$ ($\unit{1}\hour\,\unit{7}\min$), $\unit{16000}\sec$ ($\unit{4}\hour\,\unit{27}\min$) and $\unit{32000}\sec$ ($\unit{8}\hour\,\unit{53}\min$). Solid lines correspond to Gaussian functions with same variance according to equation \eqref{eq:conc_f_gaussian}.
    }
    \label{fig:fluo_c_x_t0}
\end{figure}

Fluorescein is introduced through injection site (1) and its concentration $C_f$ is measured as a function of space and time.  
As seen in figure \ref{fig:fluo_c_x_t0}\,$(a)$, which displays the time evolution of $\log(C_\mathit{f}(x,t)/\langle C_\mathit{f}\rangle)$, the 1D concentration profiles are rather homogeneous in convection cells, with strong gradients in between. 
This is consistent with the observations of \cite{solomon1988passive} who reported that after several hours the concentration becomes constant in each convection cell, the coarse-grained profile taking a Gaussian shape. 
This is also the case here as shown in figure \ref{fig:fluo_c_x_t0}\,$b$, where the scale evolution of the normalized profiles is well fitted by Gaussian functions with a maximum centered at the origin:  
\begin{equation}
F(x,t)=\frac{N_{0,f}}{\sqrt{2\pi\,\langle x^2\rangle_f(t)}}\exp\left(\frac{-x^2}{2\,\langle x^2\rangle_f(t)}\right)\,, \label{eq:conc_f_gaussian}
\end{equation}
where $\langle x^2\rangle_f(t)$ is the measured second order moment for the fluorescein, computed following equation \eqref{eq:first_second_moments_generic}. 

 \cite{solomon1988passive} (see also \citet{bib:moffatt1983,bib:pomeau1985,bib:guyon_etal1987,bib:shraiman1987}) proposed that such a Gaussian profile at large scale, where the concentration is homogenised through each cell, corresponds to a regime of effective diffusion  in which the second-order moment of the distribution evolves linearly in time at large time
\begin{equation}
\langle x^2\rangle(t)=\langle x^2\rangle(0)+2D^\star t \label{eq:x2t_D*} 
\end{equation}
with an effective diffusion coefficient $D^\star$ such that
\begin{equation}
D^\star/D= B \sqrt{\mathit{Pe}}\,,\label{eq:D_star}
\end{equation}
where $B$ is a constant that depends on the geometry of the system, $D$ is the diffusion coefficient, and $\mathit{Pe}$ the P\'eclet number.

\begin{figure}
    \centering
    \includegraphics{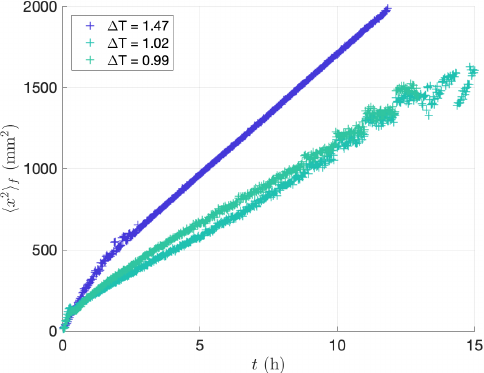}
    \caption{Typical dispersion curves for fluorescein and different values of the temperature difference $\Delta T=\{0.99;1.02;1.47\}$ (corresponding to P\'eclet numbers of the fluorescein $\mathit{Pe}_\mathit{f}$ in between $5760$ and $8630$). As found previously in\cite{solomon1988passive}, the global behavior corresponds to an effective diffusion, with an effective diffusion coefficient given by equation \eqref{eq:x^2_f}}
    \label{fig:fluo_x2_t}
\end{figure}

Typical curves of the second order moment $\langle x^2\rangle_f(t)$ for the fluorescein and different P\'eclet numbers are shown in figure \ref{fig:fluo_x2_t}.   
After a rather long transient of about $2$ hours, which corresponds to the time needed for invasion of a few cells, we observe that $\langle x^2\rangle_f$ grows linearly in time, the slope being an increasing function of the P\'eclet number. 
Combining equations \eqref{eq:x2t_D*} and \eqref{eq:D_star} and using a linear fit, we obtain:
\begin{equation}
    \langle x^2\rangle_f(t)\approx \langle x^2\rangle_f(0)+2\times B\times D_f \sqrt{\mathit{Pe}_f}\times t\, ,
   \label{eq:x^2_f}
\end{equation}
with $\sqrt{\langle x^2\rangle_f(0)}\approx12\,\milli\meter$, which corresponds to a little more than one cell invaded by the fluorescein  after injection at $t=0$, and $B=0.5$,
which is consistent with what was found by Solomon \& Gollub given their definition of the P\'eclet number \citep{solomon1988passive}.

\subsection{Case of salt (LiCl)}

Salt (LiCl) is introduced at a concentration of $20\,\milli\mole\per\liter$, a concentration low enough so that no gravity current is created due to a possible density mismatch;  it is then transported in the RB chamber in the same way as fluorescein.
It is thus possible to estimate its dispersion from the results obtained for fluorescein, provided that its diffusion coefficient is slightly different, although of the same order of magnitude (table \ref{Tab:1}). 
Therefore, we can extend the findings for flurorescein, and assume that the salt concentration also has a Gaussian coarse-grained profile over long timescales. 
We also  differentiate between cases where salt is introduced through the injection holes (1) and (2). 
We thus obtain the following coarse-grained profile $S^{(i)}$ for the salt:
\begin{equation}
    S^{(i)}(x,t)=\frac{N_{0,s}}{\sqrt{2\pi\,\langle x^2\rangle_s(t)}}\exp\left(\frac{-\,\bigl(L^{(i)} -x\bigr)^2}{2 \,\langle x^2\rangle_s(t)}\right)\, ,\label{eq:conc_s_Gaussian}
\end{equation}
where $i=$\{1,2\} stands for the injection hole through which the salt is injected, with $L^{(1)}=0$ and $L^{(2)}=L_\mathit{tot}$. 
The variance $\langle x^2\rangle_s(t)$ of salt spreading does not depend on the salt's direction of travel and is therefore identical in both cases, 
\begin{equation}
\langle x^2\rangle_s(t)\approx\langle x^2\rangle_s(0)+2\times B\times D_s \sqrt{\mathit{Pe}_s}\times t\, ,\label{eq:x2s(t)_numeric}
\end{equation}
with $\sqrt{\langle x^2\rangle_s(0)}\approx12\milli\meter$ and $B=0.5$, as for fluorescein (equation \ref{eq:x^2_f}).

%%%%%%%%%%%%%%%%%%%%%%%%%%%%%%%%%%%%%%%%%%
\subsection{Colloids: reference case}
\label{sec:reference}

\begin{figure}
%    \centering
    \includegraphics{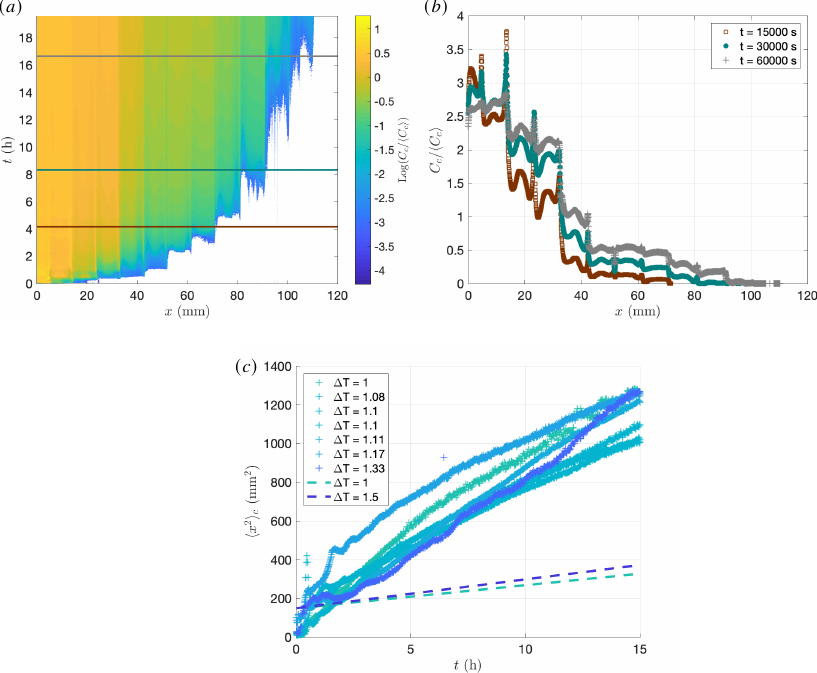}
    \caption{
    Reference case (colloids alone): $(a)$ Normalized colloid concentration profile as a function of time for $\Delta T=1.1\degreecelsius$ ($\mathit{Pe}_c=3.9\times10^5$); $(b)$: three concentration profiles extracted from figure \ref{fig:ref_x2_t}\,$a$ (corresponding to the three horizontal lines on this figure) for $t=\unit{15000}\sec$ ($\unit{4}\hour\,\unit{10}\min$), $\unit{30000}\sec$ ($\unit{8}\hour\,\unit{20}\min$) and $\unit{60000}\sec$ ($\unit{16}\hour\,\unit{40}\min$). $(c)$ Dispersion curves $\langle x^2 \rangle_c(t)$ obtained with colloids and temperature difference $\Delta T \in [1;~1.33]$ $^\circ$C. The darker the symbol, the larger the colloid P\'eclet number. 
    The two dotted lines stand for the theoretical predictions for the spreading of neutral particles of the same size as the colloids, calculated using equation 
    \eqref{eq:x^2_f} transposed to colloids, 
    $\langle x^2\rangle_c(t)\approx \langle x^2\rangle_c(0)+2\times B\times D_c \sqrt{\mathit{Pe}_c}\times t$, with $\sqrt{\langle x^2\rangle_c(0)}\approx12\milli\meter$ and $B=0.5$ as for fluorescein (equation \ref{eq:x^2_f}),
    for $\Delta T=1.\degreecelsius$ and $\Delta T=1.5\degreecelsius$.
    }
    \label{fig:ref_x2_t}
\end{figure}
 
We now investigate how colloids are transported through the chamber in the reference case, \textit{i.e.} without salt. 
Colloids are injected through site (1) at $t=0$. 
From the images, we build 1D normalized concentration profiles  $C_c(x,t)/\langle C_c(x,t) \rangle$ averaged over the height of the chamber. 
Their typical time evolution is shown in Figure \ref{fig:ref_x2_t}\,$a$ for $\Delta T=1.1\degreecelsius$ ($\mathit{Pe}_c=3.9\times10^5$). 
As clearly visible, the concentration field is much less homogeneized than in the case of fluorescein, with spatial oscillations and peaks located in between the cells even at long times. 
This is clearly visible when plotting directly the normalized concentration $C_c/\langle C_c \rangle$ at different times as displayed in \ref{fig:ref_x2_t}\,$b$. 
Such a difference between profiles measured in the case of particles and molecules in the chamber was already pointed out by \cite{solomon1988passive}, who mentioned that homogeneization does not really take place in the cells in the case of latex particles, due to their large size and hence weak diffusion coefficient. 
A first noticeable difference with results from \cite{solomon1988passive} is that the colloid concentration profiles can hardly be described as Gaussian functions, even more than 16 hours after injection. 
Another difference can be seen in Figure \ref{fig:ref_x2_t}\,$c$, which displays the time evolution of the second order moment of the colloid concentration field, $\langle x^2 \rangle_c$, for $7$ different Péclet numbers $\mathit{Pe}_c \in [1.1,~ 1.5]~10^6$. 
In none of the different runs was it possible to observe a well-defined linear evolution for the colloid concentration field, even after repeating the same experiment several times. 
An explanation for this could be the presence of an additional 3D structure to the flow which would modify mixing, making the concentration field more complex than with 2D flow. 
However, the contribution of such 3D structure was already mentioned in the work by \cite{solomon1988passive}, and was also present in the case of fluorescein for which we obtained that $\langle x^2 \rangle_f \propto t$ at long times. 
This means that we need to look for another mechanism that is present in the case of (charged) colloidal particles but not in solutes. 

\begin{figure}
    \includegraphics{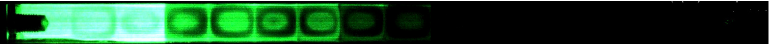}
    \caption{Typical image of a reference case ($\Delta T=1.1\degreecelsius$ and $t=43000\,\second=12\,\hour$). 
    }
    \label{fig:reference_image}
\end{figure}

Indeed, another major difference with the work by \cite{solomon1988passive} is that the effective diffusion coefficient estimated for the different values of $\Delta T$ (hence $\mathit{Pe}_c$) is much larger than expected, as can be seen from figure \ref{fig:ref_x2_t}\,$c$ in which the dotted lines represent the theoretical prediction \eqref{eq:x^2_f} transposed to neutral particles of the same size. 
At these small concentration levels, this cannot be related to a collective effect of colloids. 
Another possibility is the existence of thermophoretic effects, whereby large particles exhibit a drift velocity related to thermal gradients. 
However, this would imply that the effect increases with rising $\Delta T$, which is not the case here as can be seen in figure \ref{fig:ref_x2_t}\,$c$, where the effective diffusion is almost independent of $\Delta T$.
One can also note that the top-bottom symmetry of the observed pattern in Figure \ref{fig:reference_image}, which displays the snapshot of a typical reference case, definitely contradicts the thermophoresis hypothesis.
More likely, the explanation may come from the fact that, because we need to study diffusiophoretic effects, our particles are (positively) charged, and hence may interact with the upper and lower walls made of brass, which takes a negative charge in water \citep{sarver2016,CUI2024105291}. 
This additional mechanism tends to facilitate the diffusion of particles along the separatrixes of the flow, leading to a much larger dispersion. 
This hypothesis is reinforced when looking at Figure \ref{fig:reference_image}, where we clearly see that colloids are attracted to the walls; this effect was not present in \cite{solomon1988passive}.

%\vskip1cm

\section{Diffusiophoresis and cellular flow} \label{section:diffusio}
In this section we first summarise the theoretical results obtained in \cite{bib:volk_etal2022phoresis}, in order to transpose them to the experiment afterwards. 
In particular, we want to check whether our experimental conditions could lead to the regime of enhanced transport, or else to blockage.

%%%

\subsection{Theory from \cite{bib:volk_etal2022phoresis}}
\label{sec:theory}
\begin{figure}
	\includegraphics{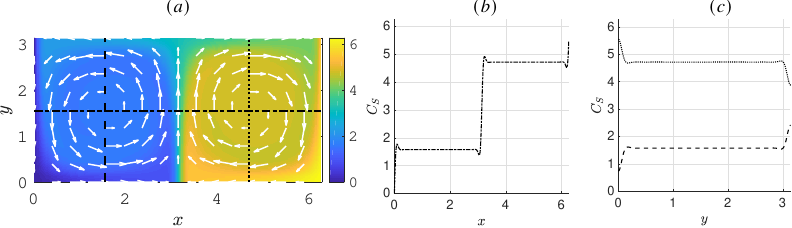}
    \caption{Salt distribution (numerical simulations) in the stationary regime, for a constant horizontal mean gradient $\mathbf{G}=G\,\mathbf{e}_x$; here $\mathit{Pe_s}=3140$, $G=1$ and the size of a cell is $L=\pi$. $(a)$: stationary salt concentration field $C_s$ in two consecutive cells, with an imposed constant mean gradient $\mathbf{G}=G\,\mathbf{e}_x$; the white arrows represent the cellular velocity field. $(b)$: horizontal profile along the horizontal dashed line in subfigure $a$; $(c)$: 2 vertical profiles along the vertical dashed lines. The salt concentration is roughly constant in the cells, except around the vertical and horizontal separatrixes. 
    }
    \label{fig:salt_profiles_theory}
\end{figure}

%%%%%
The theoretical results in \cite{bib:volk_etal2022phoresis} were obtained in \textit{stationary} conditions. 
The advection-diffusion equation was first solved numerically for the salt concentration, $C_s$, in the presence of an imposed mean salt gradient, $\mathbf{G}$, using an analytical cellular velocity field, $\mathbf{u}$. 
In the stationary regime, this results in a solution where the salt concentration is almost constant within each cell, except in the vicinity of the separatrixes, see figure \ref{fig:salt_profiles_theory}\,$(a)$ for an example with $\mathit{Pe}_s=1340$. 
The horizontal profile shown in figure \ref{fig:salt_profiles_theory}\,$(b)$ exhibits a sharp increase of magnitude $\Delta S=G\times L$ between two consecutive cells, indicating a very high salt gradient. 
The width $\ell_s$ of this gradient is found by the competition between contraction by the flow (typical time $L/U$) and diffusion (typical time $\ell_s^2/D_s$), hence $\ell_s\sim L/\sqrt{\mathit{Pe}_s}$, and the corresponding salt gradient in this region is $\Delta S/\ell_s\sim G\sqrt{\mathit{Pe}_s}\gg G$. 
Note that the vertical profiles in figure \ref{fig:salt_profiles_theory}\,$(c)$ also show a small vertical salt gradient around the horizontal separatrixes, but of much weaker intensity; hence we neglect it in a first step.

\begin{figure}
    \includegraphics{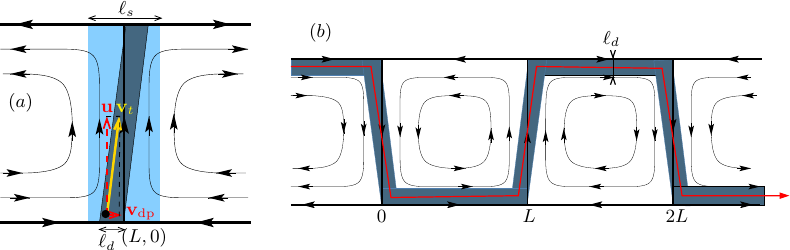}
    \caption{$(a)$ Schematic displacement of a particle in the salt gradient (light blue region) around the vertical ascending separatrix $x=L$.
The Eulerian velocity field $\mathbf{u}$ is vertical and ascending, and $\mathbf{v}_\mathrm{dp}$ is horizontal; 
the total velocity of particles around the vertical separatrixes is therefore  $\mathbf{v}_t=\mathbf{u}+\mathbf{v}_\mathrm{dp}$, with $\mathbf{u}\approx U\, \mathbf{e}_y$. 
Particles that change cell are located on the grey strip of width $\ell_d$. 
$(b)$ Schematic location of particles that change cell, in grey, for cells located between $x=0$ and $x=2L$; 
an example of a trajectory is shown in red. 
Outside this grey strip, the particles remain trapped in their cells. From \cite{bib:volk_etal2022phoresis}.
}    \label{fig:colloids_displacement_theory}
\end{figure}
We now consider the displacement of colloids in this salt distribution. 
In \cite{bib:volk_etal2022phoresis}, the drift velocity of the active particles is of the type $\mathbf{v}_\mathrm{dp}=\alpha \nabla C_s$, where $\alpha$ is the phoretic coefficient. 
The total velocity of a phoretic particle is thus $\mathbf{v}=\mathbf{u}+\mathbf{v}_\mathrm{dp}$, where $v_\mathrm{dp}\ll U$. 
When a particle enters the salt gradient (shown in light blue in figure \ref{fig:colloids_displacement_theory}\,$(a)$), it deviates horizontally by a distance $\ell_d$ such that $\ell_d/L=v_\mathrm{dp}/U$, so that $\ell_d\sim\alpha G L \sqrt{\mathit{Pe}_s}/U$. 
The particles that change cell are located within the grey strip of width $\ell_d$ in figure \ref{fig:colloids_displacement_theory}\,$(a)$. 
These particles are then transported horizontally by the velocity field until they reach the next region of high salt gradient, thus travelling around the cells, globally in the direction of the mean salt gradient $\mathbf{G}$, as depicted in figure \ref{fig:colloids_displacement_theory}\,$(b)$. 
As the particles located outside the grey strip do not change cell, it is possible to estimate the mean global horizontal velocity as 
$v_m\sim 1/{L^2}\times\left[(\ell_d\times L)\,v_\mathrm{dp}+(\ell_d\times L)\,U\right]$. 
Finally, because $v_\mathrm{dp}\ll U$, we obtain 
\begin{equation}
   v_m \sim \alpha G \sqrt{\mathit{Pe}_s}\,.
   \label{eq:Vm_theory}
\end{equation}
This regime, for which the mean velocity is much higher than the velocity, $\alpha G$, associated to the mean salt gradient, is called ``enhanced transport''. 
Lagrangian numerical simulations performed in \cite{bib:volk_etal2022phoresis} show that equation \eqref{eq:Vm_theory} gives a good estimation of the global mean velocity. 

The blockage regime is related to the weaker vertical salt gradients around the horizontal separatrixes, see figure \ref{fig:salt_profiles_theory}\,$(c)$.  
Indeed, because the salt concentration increases from left to right due to the imposed mean gradient $\mathbf{G}$, the velocity field transports a deficit of salt when travelling horizontally from left to right, and an excess of salt when travelling in the opposite direction, as visible in figure \ref{fig:salt_profiles_theory}\,$(a)$. 
Therefore, the phoretic particles located within a horizontal segment of the grey strip in figure \ref{fig:colloids_displacement_theory}\,$(b)$ always encounter a salt gradient pointing vertically into the cell, and gradually deviate towards its centre:
after some time, they leave the grey strip for the central region of the cell. 
Once all the particles have left the strip, they cannot change cell anymore and become blocked.
In addition, any fluid particle located inside the strip follows the velocity field and completes one revolution around the cell in a typical time of $4L/U$. 
This creates a corona of width $\ell_d$ devoid of particles around each cell, as demonstrated numerically in \cite{bib:volk_etal2022phoresis} (see also Fig. \ref{fig:blockage_image} for an experimental visualisation).

This blocking mechanism only occurs when diffusion is insufficient to bring particles back into the strip. 
Therefore, we need to estimate how long it takes for particles outside the strip to re-enter it. 
Consider a strip of width $\ell_d$ empty of particles in a medium that is filled with particles. 
Due to the linearity of the advection-diffusion equation, the problem is the same as considering a strip of width $\ell_d$ filled with particles in an empty medium: the typical diffusion time is, in both cases, $\ell_d^2/D_c$.

Blockage therefore corresponds to a situation in which fluid in the empty corona can travel around the cell while remaining empty of particles, which writes
$4L/U \ll \ell_d^2/D_c$. 
Introducing the blockage parameter $R$ such that
\begin{equation}
    R =\frac{\alpha\, GL}{2\sqrt{D_c D_s}}\,, 
\label{eq:blockage_coeff}
\end{equation}
we obtain the blockage situation as corresponding to $R\gg1$. 

Numerical simulations performed in \cite{bib:volk_etal2022phoresis} demonstrate a sharp transition between blockage and enhanced transport, depending on $R$: 
a complete blockage is observed whenever $R\ge 4$, while enhanced transport, with a magnitude given by equation \eqref{eq:Vm_theory}, corresponds to $R\le 1$; in practice enhanced transport is already observed for $R\sim 2$, although with a weaker velocity. 
This is all explained in much more detail in \cite{bib:volk_etal2022phoresis}.

\subsection{Experiment versus theory}
Although there are some differences between the experiment and the theory, we now demonstrate how the theoretical findings can be adapted to the experimental conditions. 

\subsubsection{Velocity field}
In the theory, numerical work and experiments, the velocity field is steady and cellular. 
There are some minor discrepancies, such as the boundary conditions applied to the upper and lower walls (no-slip in the experiment, and slip in the numerical analysis), but it should be noted that the theory is not heavily dependent on the specifics of the flow. 
On the one hand, the enhanced transport regime only depends on the salt P\'eclet number $\mathit{Pe}_s$, based on the maximum velocity $U$,  as shown in equation  \eqref{eq:Vm_theory}. 
On the other hand, the blockage parameter $R$ (equation \ref{eq:blockage_coeff}) does not depend on the velocity at all. 
Therefore, we can conclude that the theory and the experiment are very similar in this respect. 

Finally the major difference between the theory and the experiment is that the former considers an infinite medium, whereas the latter has a finite extent. 
This is not important when considering the blockage situation, but it will be important when considering enhanced dispersion. 

%%%
\subsubsection{Salt distribution} 
The situation regarding the distribution of salt is very different: the theory assumes an \textit{imposed mean salt gradient} which is assumed to be constant in space and time. Furthermore, the theory deals with the \textit{stationary solution} of the advection-diffusion equation. 
By contrast, the salt distribution in the experiment is time-dependent and follows a Gaussian coarse-grained profile, denoted by $S$ (see equation \ref{eq:conc_s_Gaussian}).
The experimental equivalent for $G$ is the horizontal salt concentration jump across a cell, divided by the cell's length $L$. 
This jump in salt concentration can be estimated rather accurately using the coarse-grained profile $S$. 
If we denote as $G^{(i)}$ the mean horizontal salt gradient across a cell in the case when salt is introduced through site $(i)$, then $G^{(i)}\approx dS/dx$, so that we obtain from equation \eqref{eq:conc_s_Gaussian}:
\begin{equation}
 G^{(i)}(x,t)=\frac{x-L^{(i)}}{\langle x^2\rangle_s(t)}\times S(x,t) \, ,
\end{equation}
where we remind that $L^{(1)}=0$ and $L^{(2)}=L_\mathit{tot}$.

\subsubsection{Velocity drift} 
As stated in paragraph \ref{sec:theory}, in \cite{bib:volk_etal2022phoresis} the drift velocity due to phoresis was taken as $\mathbf{v}_\mathrm{dp}=\alpha \nabla C_s$. 
However, the relationship between drift velocity and scalar gradients is generally much more complex \citep{bib:gupta_etal2020, bib:lee_etal_2023}. 
For very charged particles such as ours, the drift velocity takes the form \citep{bib:ault2024}:
\begin{equation}
    \mathbf{v}_\mathrm{dp}= D_\mathrm{dp} \nabla \log C_s= D_\mathrm{dp} \frac{\nabla C_s}{C_s}\,.
    \label{eq:v_dp}
\end{equation}
Then we switch from theory to experiment by setting $\alpha\approx D_\mathrm{dp}/C_s$, or rather $\alpha\approx D_\mathrm{dp}/S$ when dealing with mean quantities as in equations \eqref{eq:Vm_theory} or \eqref{eq:blockage_coeff}. 
Note also that the experiments may be non-repeatable in the case of salt concentrations $C_s$ that are too small, far from the injection hole at early times. This will be taken into account later. 

%%%
\subsubsection{Enhanced transport}
Even if the theoretical formula \eqref{eq:Vm_theory} and \eqref{eq:blockage_coeff} were derived in a stationary regime, they may still apply to a transient case, provided that the salt evolves slowly over time, as is the case in the experiment. 
From the analysis above, we can transpose the value found for enhanced transport in equation \eqref{eq:Vm_theory} to the experimental conditions. We obtain
   \begin{equation}
    v_m^{(i)}(x,t)\sim D_\mathrm{dp} \sqrt{\mathit{Pe}_s} \,\frac{L^{(i)}-x}{\langle x^2\rangle_s(t)} \, ,
    \label{eq:Vm^i_x_theo}
    \end{equation}
where again $i=\{1,2\}$ stands for the injection site. 
We observe that the velocity due to salt attraction is also inhomogeneous and time-dependent.  
When $i=1$ ($L^{(1)}=0$), the velocity is negative, as expected for colloids that move in the direction of the salt gradient; conversely, when $i=2$ ($L^{(2)}=L_\mathit{tot}$), the velocity is positive. 
Note also that the effect decays over time, but increases with the P\'eclet number.

\subsubsection{Blockage coefficient}
We now transpose the theoretical blockage coefficient (equation \ref{eq:blockage_coeff}) to experimental conditions. 
Unlike enhanced transport, blockage does not depend on the direction of the salt gradient. 
Therefore the blockage coefficient is always positive and is given by:
\begin{equation}
    R^{(i)}(x,t)=\frac{|L^{(i)}-x|\times L}{2\,\langle x^2\rangle_s(t)}\times\frac{D_\mathrm{dp}}{\sqrt{D_cD_s}}\,. \label{eq:coeff_blockage_R_i}
\end{equation}
In the experiment, we have $D_s =1360$ $\mu\mathrm{m}^2 \mathrm{s}^{-1}$, $D_c = 2$ $\mu\mathrm{m}^2 \mathrm{s}^{-1}$ and $D_\mathrm{dp} = 290 \mu\mathrm{m}^2 \mathrm{s}^{-1}$, so that $D_\mathrm{dp}/\sqrt{D_cD_s}\simeq 5.56$. 
Because the theory predicts blockage for $R\ge 4$, blockage may be possible to observe at some location. 
Note also that $\langle x^2\rangle_s(t)$ increases with time, implying that the blockage coefficient $R$ decreases with time at a given point $x$, so that  blockage is not possible at long time. 

However, the value of the blockage coefficient cannot be transposed directly to the experiment status. 
Indeed, the theory was derived in the context of an \textit{infinite and periodic} medium, and in a \textit{stationary} state: the conditions in the experiment differ greatly from these.

\begin{table}
\begin{center}
\begin{tabular}{|Sc|Sc|Sc|}
    \Xhline{0.5pt}
         & Theory  & Experiment\\
\Xhline{0.5pt}         
$ \text{velocity field}$  & cellullar and steady  & cellullar and steady\\
&infinite medium& finite extent\\
\Xhline{0.5pt}
 &  stationary &  non stationary\\
salt distribution & imposed constant mean gradient $G$& mean gradient $G^{(i)}(x,t)=\displaystyle\frac{x-L^{(i)}}{\langle x^2\rangle_s(t)}\times S(x,t)$ \\
\Xhline{0.5pt}
drift velocity & $\mathbf{v}_\mathrm{dp}=\alpha \nabla C_s$ & $\mathbf{v}_\mathrm{dp}=D_\mathrm{dp} \nabla \log C_s=D_\mathrm{dp} \displaystyle\frac{\nabla C_s}{C_s} $\\
\Xhline{0.5pt}
enhanced transport &  $v_m \sim \alpha G \sqrt{\mathit{Pe}_s}$  & $v_m^{(i)}(x,t)\sim D_\mathrm{dp} \sqrt{\mathit{Pe}_s} \,\displaystyle\frac{L^{(i)}-x}{\langle x^2\rangle_s(t)}$\\
\Xhline{0.5pt}
blockage coefficient & $R =\displaystyle\frac{\alpha\, GL}{2\sqrt{D_c D_s}}$ & $R^{(i)}(x,t)=\displaystyle\frac{|L^{(i)}-x|\times L}{2\,\langle x^2\rangle_s(t)}\times\frac{D_\mathrm{dp}}{\sqrt{D_cD_s}}$\\
\Xhline{0.5pt}
\end{tabular}
\end{center}
   \caption{Summary of the differences between theory and experimental conditions. For the experiment, the index $i=\{1,2\}$ stands for the injection hole of the salt (figure \ref{fig:cell_setup}), with $L^{(1)}=0$ and $L^{(2)}=L_\mathit{tot}$. }   \label{tab:theory_vs_exp}
\end{table}

The differences between theory and experiment are summarized in table \ref{tab:theory_vs_exp}. 
In the following we explain the consequences for both  the enhanced transport and blockage configurations. 
%\\

%\newpage
%%%%
\subsection{Salt-out configuration: enhanced transport}
In practice, blockage or enhanced transport could, in theory, be observed regardless the configuration considered (salt-in or salt-out). 
Indeed, let us come back to the value of the blockage coefficient. 
In the salt-out configuration, it is obtained by letting $i={\scriptsize 2}$ in equation \eqref{eq:coeff_blockage_R_i}:
\begin{equation}
    R^\mathit{salt-out}(x,t)=\frac{(L_\mathit{tot}-x)\times L}{2\,\langle x^2\rangle_s(t)}\times\frac{D_\mathrm{dp}}{\sqrt{D_cD_s}}\,. \label{eq:coeff_blockage_R_salt-out}  
\end{equation}
\unitlength=1.mm
\begin{figure}
    \centering
    \includegraphics{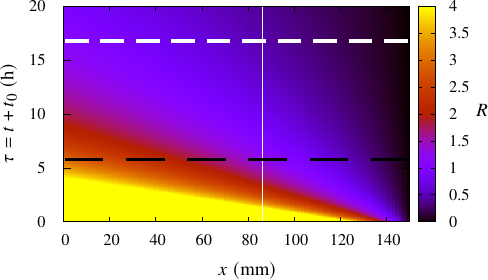}   
    \caption{Space-time diagram for the blockage coefficient $R$ in the salt-out configuration (equation \ref{eq:coeff_blockage_R_salt-out}) for $\Delta T=1\degreecelsius$, with the experimental data $D_s =1360$ $\mu\mathrm{m}^2 \mathrm{s}^{-1}$, $D_c = 2$ $\mu\mathrm{m}^2 \mathrm{s}^{-1}$ and $D_\mathrm{dp} = 290\, \mu\mathrm{m}^2 \mathrm{s}^{-1}$ ($D_\mathrm{dp}/\sqrt{D_cD_s}\simeq 5.56$). 
    Values higher than $4$, which correspond to possible blockage, are printed in yellow. The salt is released at time $\tau=0$, while the colloids are released a time $t_0$ later; we set $t=\tau-t_0$, so as to have the colloids released at $t=0$. The dashed lines indicate the time at which the colloids are released in the experiments presented below: black dashed line: $t_0=6\,\hour$; white dashed lines: $t_0=17\,\hour$.}
    \label{fig:R_salt_out}
\end{figure}
The space-time diagram for $R^\mathit{salt-out}$ is shown in figure \ref{fig:R_salt_out}; values higher than $4$, corresponding to possible blockage, are coloured yellow.
The diagram shows high values of $R$ (in yellow) at short times in the region where the colloids are injected and initially spread, which is consistent with possible blockage.
However, blockage is not physically reasonable: 
actually, as previously explained, the salt Gaussian coarse-grained profile used in the calculation does not settle instantaneously in the chamber. 
Furthermore, it takes time for the salt to travel across the whole cell towards the colloids. 
Finally, note that these large values of $R$ would correspond to very small salt concentrations here, which could lead to non-repeatable experiments (see equation \ref{eq:v_dp}). 
Therefore, the salt must be injected into the cell much before the colloids, so as to have reached the injection site (1) when the colloids are introduced: as shown in figure \ref{fig:cell_setup_salt-out}, salt is introduced at time $t=-t_0$, with $t_0>0$ the time-lag before colloids injection. 
We let $\tau=t+t_0$; salt is therefore introduced at $\tau=0$ in this time coordinate, and $\langle x^2\rangle_s (\tau)$ verifies exactly equation \eqref{eq:x2s(t)_numeric}. 
We can write: 
 \begin{equation}
   \langle x^2\rangle_s(\tau)=\langle x^2\rangle_s(t+t_0)\approx  \langle x^2\rangle_s(0)+2\times B\times D_s\sqrt{\mathit{Pe}_s} \times (t+t_0)\, , \label{eq:x2s_saltout_t0_num}
\end{equation}
with $\sqrt{\langle x^2\rangle_s(0)}\approx 12\,\milli\meter$ and $B\approx 0.5$  from equation \eqref{eq:x2s(t)_numeric}. 
\begin{figure}
    \centering
    \includegraphics{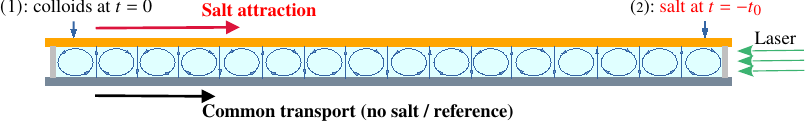}
    \caption{
    Salt-out configuration: first, salt (LiCl, non fluorescent) is introduced through injection site (2) at $t=-t_0$; then, colloids (fluorescent) are introduced through injection site (1) at $t=0$. The delay $t_0$ must be long enough so that some salt has reached the other side of the cell when colloids are introduced. The two horizontal bold arrows show the transient effects felt by the colloids after introduction: they naturally travel towards the right, as showed in the reference case (common transport); they are also attracted by salt and therefore furthermore travel toward the right (salt attraction). 
    }
    \label{fig:cell_setup_salt-out}
\end{figure}

In practice, we found no noticeable diffusiophoretic effect of salt on the colloids whenever the time-lag $t_0$ was shorter than $5$ to $6$ hours. 
In figure \ref{fig:R_salt_out}, the line corresponding to the time-lag $t_0=6\,\hour$, corresponding to one of our experiments, is plotted in black dotted line: as can be seen, at that time the blockage coefficient $R$ is already too small to have blockage. 
Furthermore, all transient effects, (\textit{i.e.} the common transport of colloids and salt attraction) tend to move the colloids to the right (figure \ref{fig:cell_setup_salt-out}), which works against blockage. 
Finally, note that $R$ decreases from left to right, in the direction of colloid movement.
Therefore, the salt-out configuration is highly unlikely to exhibit blockage, but it may result in enhanced transport. 

We now apply equation \ref{eq:Vm^i_x_theo} with $i=2$ ($L^{(2)}=L_\mathit{tot})$ and we get:  
\begin{equation}
    v_m(x,t)\sim D_\mathrm{dp} \sqrt{\mathit{Pe}_s} \,\frac{L_{tot}-x}{\langle x^2\rangle_s(t+t_0)} \, ,
    \label{Vm_x_theo_saltout}
    \end{equation}
with $\langle x^2\rangle_s$ given by equation \eqref{eq:x2s_saltout_t0_num}.
This equation can be averaged over the colloid concentration profile to obtain the global mean velocity of colloids in the chamber $V_m(t)=\langle v_m(x,t)\rangle_c$:
\begin{equation}
V_m(t)=\frac{1}{L_{tot}\,\langle C\rangle}\int_0^{L_{tot}} C(x)\, v_m(x)\,dx\sim\frac{D_\mathrm{dp}\sqrt{\mathit{Pe}_s}}{\langle x^2\rangle_s(t+t_0)}\,\bigl(L_{tot}-\langle x\rangle_c(t)\bigr). \label{eq:Vm(t)}
\end{equation}
Using the equality:
\begin{equation}
V_m(t)=\frac{d\langle x\rangle_c(t)}{dt} \label{eq:v_m(t)dxcdt}\,,
\end{equation}
it is possible to predict the evolution of $\langle x\rangle_c(t)$. 
Indeed, combining equations \eqref{eq:Vm(t)} and \eqref{eq:v_m(t)dxcdt}, we obtain
\begin{equation}
\frac{d\langle x\rangle_c}{L_{tot}-\langle x\rangle_c(t)}= D_\mathrm{dp} \sqrt{\mathit{Pe}_s}\,\frac{dt}{\langle x^2\rangle_s(t+t_0)}
\end{equation}
with $\langle x^2\rangle_s(t+t_0) =a+b(t_0+t)$ (equation \ref{eq:x2s_saltout_t0_num}).  
This can therefore be integrated into:
\begin{equation}
-\ln\frac{L_{tot}-\langle x\rangle_c(t)}{L_{tot}-\langle x\rangle_c(0)}=D_\mathrm{dp} \sqrt{\mathit{Pe}_s}\times\frac{1}{b}\,\ln\frac{a+b(t_0+t)}{a+b\,t_0}\,.
\end{equation}
Introducing the non-dimensional coefficient $\beta$:
\begin{equation}
\beta=D_\mathrm{dp} \sqrt{\mathit{Pe}_s}\times\frac{1}{b}=\,\frac{D_\mathrm{dp}\sqrt{ \mathit{Pe}_s}}{2D^\star_s}, \label{eq:beta}
\end{equation}
which verifies $\beta\sim\frac{D_\mathrm{dp}}{D_s}<1$ as $D^\star_s\sim 0.5\, D_s\sqrt{\mathit{Pe}_s}$,
we finally obtain:
\begin{eqnarray}
\langle x\rangle_c(t)&=&L_{tot}-\Bigl(L_{tot}-\langle x\rangle_c(0)\Bigr)\left(\frac{a+b\,t_0}{a+b\,(t_0+t)}\right)^\beta \label{eq:xc_theorique}\\
&=&L_{tot}-\Bigl(L_{tot}-\langle x\rangle_c(0)\Bigr)\left(\frac{\langle x^2\rangle_s(t_0)}{\langle x^2\rangle_s(t_0+t)}\right)^\beta \label{eq:x_c_salt_out_final}
\end{eqnarray}
with  $a$ and $b$ given in equation \eqref{eq:x2s(t)_numeric} and $\beta$ given by equation \eqref{eq:beta}. 

The formula implies that as $t\rightarrow\infty$, the ratio of salt dispersion becomes small, and $\langle x\rangle_c(t)$ tends to $L_{tot}$; 
physically one would rather think of $L_{tot}/2$. 
But of course there is no diffusion in those calculations, only physics of particles which move with a velocity proportional to $\nabla S$, which only cancels at $x=L_{tot}$. 
Similarly, at short time we have
\begin{equation}
\langle x\rangle_c(t)\underset{t\rightarrow0}{\sim}L_{tot}-\Bigl(L_{tot}-\langle x\rangle_c(0)\Bigr)\times \left(1- \frac{D_\mathrm{dp}\sqrt{\mathit{Pe}_s}}{\langle x^2\rangle_s(t_0)} t\right)\, ,
\end{equation}
so that the slope at the origin is all the more important as $t_0$ is small. 
As previously explained, this is, of course, not valid if colloids are introduced too early after salt injection ($t_0\le 5\,\hour$), since the calculation assumes that the salt Gaussian profile has been established throughout the cell.  

\begin{figure}
    \centering
    \includegraphics{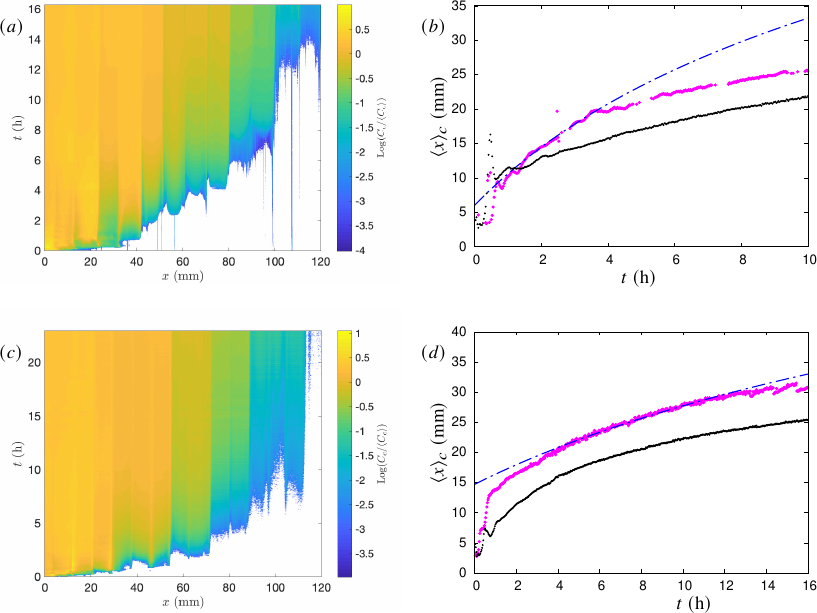}
   \caption{Salt-out case. left: colloid concentration profile as a function of time for 2 different time-lags $t_0$; right: corresponding $\langle x\rangle_c$ as a function of time (magenta curve). The dotted blue line is the theoretical prediction (equation \ref{eq:x_c_salt_out_final}). In view of comparison, the black curve is a reference case (colloids with no salt) for a similar P\'eclet number. 
   Top: $t_0=\unit{6}\hour$ ($\Delta T=1.03\degreecelsius$);  bottom: $t_0=\unit{17}\hour$ ($\Delta T=1.1\degreecelsius$).}
    \label{fig:saltout_cxt}
\end{figure}
%%%
The experimental results in the salt-out configuration are shown in figure \ref{fig:saltout_cxt}, for two different values of the time-lag $t_0$, and rather similar P\'eclet numbers ($\Delta T=1.03\degreecelsius$ and $\Delta T=1.1\degreecelsius$).  
On the right, the magenta curves show the mean longitudinal profile $\langle x\rangle_c(t)$; 
in view of comparison, the black curves show the same quantity in a reference case (colloids with no salt).
First of all, one can notice that the effect of salt compared to the reference, although visible, is not as important as supposed from the theory. 
Indeed, as explained before, due to their charge, the colloids spread much more rapidly in the cell than what is expected for non-charged particles (see also figure \ref{fig:ref_x2_t}\,$(c)$). 
The theoretical prediction for the salt effect (equation \ref{eq:x_c_salt_out_final}) is plotted in dotted blue line for the two time-lags. 
As can be seen, the order of magnitude of the salt effect is rather well predicted by equation \eqref{eq:x_c_salt_out_final}. 
Indeed, as shown in figure \ref{fig:R_salt_out}, these two time-lags correspond to moments when the blockage coefficient is far enough from blockage conditions so as to allow enhanced transport. 
This is completely achieved for $t_0=17\hour$ (white dotted line), but also true for $t_0=6\hour$ (black dotted line) provided that $x\ge10\,\milli\meter$, \textit{i.e.} from the second cell onwards.

After some time, the experimental profile does not keep up with the prediction anymore, but rather follows the same slope (hence same velocity) as the reference. 
This can be explained easily from the fact that the experiment has a finite size: 
indeed, in the theory, the regime of enhanced transport derives from a strip of colloids which travels in the direction of the mean salt gradient from left to right, see Figure \ref{fig:colloids_displacement_theory}. 
In the theory, the domain is infinite, so that there are always colloids arriving from the left. 
Because the experiment is finite, after a while the strip on the left has emptied and cannot be renewed, which stops the process. 
We can estimate the time spent by the colloids in the strip to cross the whole cell. 
The time needed to change cell is $\ell_d/v_\mathrm{dp}\sim L/U$, see figure \ref{fig:colloids_displacement_theory}\,$(a)$. Since there are $N$ cells in the chamber, the total time spent by particles to cross all the vertical gradients is $(N-1)\times L/U\approx L_\mathit{tot}/U$. 
Outside the salt gradient, the colloids are transported by the flow, see figure \ref{fig:colloids_displacement_theory}\,$(b)$. 
Therefore the time spent by a colloid to cross a cell from one vertical salt-gradient to the other is also $L/U$, with a total time in the horizontal strips of the order of $L_\mathit{tot}/U$. 
Finally, the total time spent by the colloids initially in the left strip to cross the entire cell is $\tau_\mathit{cell}\sim 2\times L_\mathit{tot}/U$. 
With a maximum velocity $U\sim 0.3\,\milli\meter\per\second$, we obtain for our experiment $\tau_\mathit{cell}\sim 15\,\minute$. 
This result must be taken with caution: of course, the velocity near the top and bottom walls may be smaller than the maximum velocity $U$. But, more  problematically, the velocity $v_\mathrm{dp}$ in the salt gradient is very small; therefore a small error/change in $v_\mathrm{dp}$ may lead to a quite different result. 
For instance, in figure \ref{fig:saltout_cxt}\,$(b)$ ($t_0=6\,\hour$), we observe that, once settled, the experimental profile follows the theory of enhanced transport during two to three hours, before following the same slope as the reference (common transport). 
Although this lapse of time is rather higher than what is predicted by the theory, its order of magnitude is still reasonable, which validates the hypothesis. 
In the case $t_0=17\,\hour$ (figure \ref{fig:saltout_cxt}\,$(d)$), there is no clear transition as in the preceding case, since the slope predicted by the theory is only slighly higher than what is observed without salt.

%\newpage
%%%%%%
\subsection{Salt-in configuration: blockage}
\begin{figure}
%    \centering
	\includegraphics{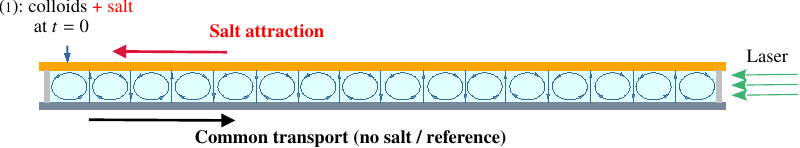}    
\caption{
Salt-in configuration: fluorescent colloids, together with non fluorescent LiCl salt,  are introduced through injection site (1) at $t=0$. The two horizontal bold arrows show the transient effects felt by the colloids after introduction: they naturally tend to travel towards the right, as showed in the reference case (common transport); they are also attracted by salt and therefore also tend to travel toward the left. There is thus a competition at small time between both processes. 
    }
    \label{fig:cell_setup_salt-in}
\end{figure}

We now turn to the salt-in configuration, depicted in figure \ref{fig:cell_setup_salt-in}:
fluorescent colloids are introduced together with non-fluorescent LiCl salt at $t=0$ through injection site(1). 
As before, we track the fluorescent species so as to measure the local concentration at each point. 

Because salt travels more rapidly than the colloidal particles, one can assume that the Gaussian salt profile seen by the colloids is rapidly settled from the injection point (1). 
In this case, the blockage coefficient reads, from equation \eqref{eq:coeff_blockage_R_i} with $i=1$ and $L^{(1)}=0$:
\begin{equation}
    R^\mathit{salt-in}(x,t)=\frac{x\times L}{2\,\langle x^2\rangle_s(t)}\times\frac{D_\mathrm{dp}}{\sqrt{D_cD_s}}\,, \label{eq:coeff_blockage_R_in}
\end{equation}
with $\langle x^2\rangle_s(t)$ given by equation \eqref{eq:x2s(t)_numeric}. \unitlength=1.mm
\begin{figure}
    \centering
   \includegraphics{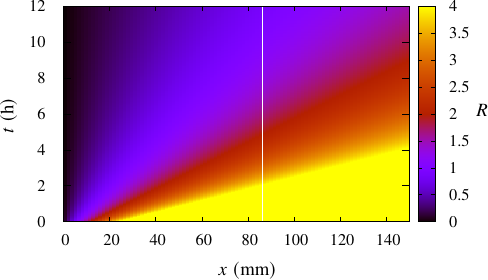} 
    \caption{Space-time diagram for the blockage coefficient $R$ in the salt-in configuration for $\Delta T=1\degreecelsius$. Values higher than $4$, which correspond to possible blockage, are printed in yellow. Both salt and colloids are released at time $t=0$ (figure \ref{fig:cell_setup_salt-in}). 
}
    \label{fig:R_salt_in}
\end{figure}
Figure \ref{fig:R_salt_in} displays the space diagram for the blockage coefficient in the salt-in configuration, for $\Delta T=1\degreecelsius$. 
It shows that blockage is possible over a short period of time once the colloids have travelled far enough away from the injection site (1); in practice, a few cells is enough. 
However, this coefficient was derived under stationary conditions. 
In practice, blockage does not occur instantly, and the transient effects that take place in the cell before blockage may happen must be taken into account. 
The first of these, as seen in figure \ref{fig:cell_setup_salt-in}, is again the common transport which takes place even without salt and tends to move the colloids to the right. 
Although one could then think that it should help blockage (the blockage coefficient is larger further from the injection site), in fact it prevents blockage since colloids that change cell are those located near the walls (figure \ref{fig:reference_image}). 
Colloids which get blocked must have remained long enough in their cell, far from the wall. 
The second transient effect is still related to salt attraction, and in those conditions makes colloids located near the walls (figure \ref{fig:colloids_displacement_theory}) travel to the left. 
Therefore the two effects push in opposite directions so that blockage is more likely to be observed. 
The condition $R\ge4$ implies, from equation \eqref{eq:coeff_blockage_R_in}:
\begin{equation}
\frac{x}{L}\ge 8\,\frac{\langle x^2\rangle_s(t)}{L^2}\,\frac{\sqrt{D_cD_s}}{D_\mathrm{dp}},\label{eq:cond_block_x}
\end{equation}
where equation \eqref{eq:cond_block_x} is therefore a necessary condition of blockage.

%--------------------------
\begin{figure}
    \includegraphics{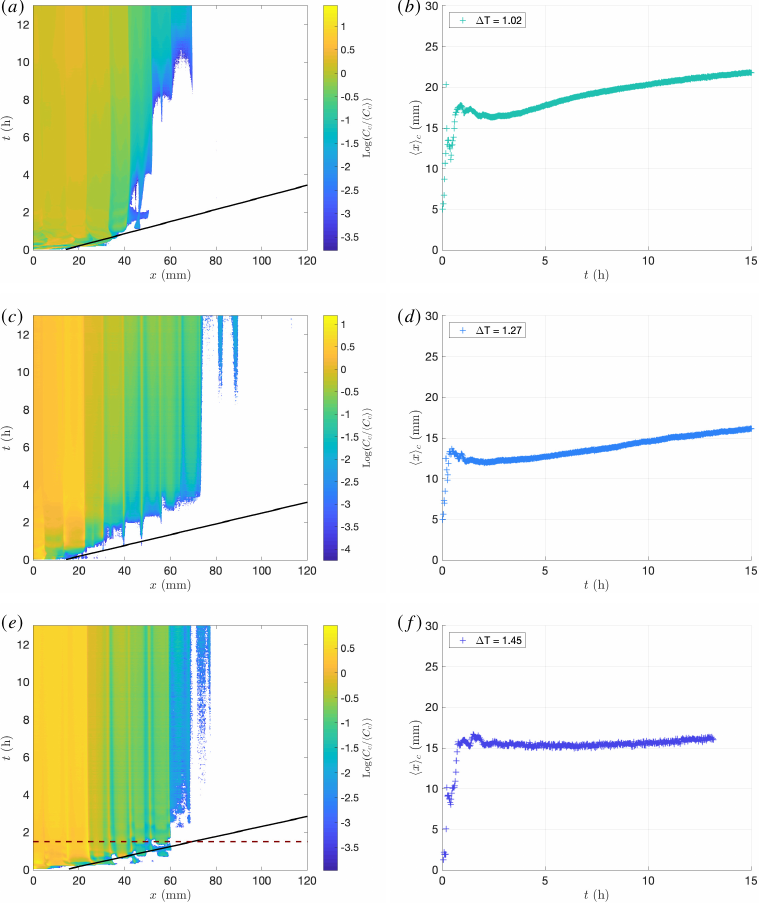}
    \caption{Salt-in configuration: Normalized concentration profiles in $x$ (left column), together with their average position $\langle x\rangle_c(t)$ (right column), for three different values of the temperature difference $\Delta T$, increasing from top to bottom. Top (subfigures \ref{fig:blockage_x}\,$a$ and $b$: $\Delta T=1.02\degreecelsius$; middle (subfigures \ref{fig:blockage_x}\,$c$ and $d$: $\Delta T=1.26\degreecelsius$; bottom (subfigures \ref{fig:blockage_x} $e$ and $f$: $\Delta T=1.45\degreecelsius$.
    The solid line has equation $x=1.43 \,\langle x^2 \rangle_s(t)/L$ (from condition \ref{eq:cond_block_x}), with $L=9.4\,\milli\meter$ and $\langle x^2 \rangle_s(t) \approx 9.4^2 + 13 \cdot 10^{-4} \sqrt{\mathit{Pe}_s} t$ (equation \ref{eq:x2s(t)_numeric}). 
    Blockage may happen whenever the profile crosses this line (necessary condition). From the mean profiles $\langle x^2\rangle_c(t)$ on the right, it is visible that blockage lasts the most for the largest temperature difference $\Delta T$ (largest P\'eclet number). 
    The dashed line in subfigure \ref{fig:blockage_x}\,$e$ at time $t=1.5\,\hour$ is the time corresponding to the snapshot of figure \ref{fig:blockage_image}.
    }
    \label{fig:blockage_x}
\end{figure}

The concentration profiles in $x$ are shown in Figure \ref{fig:blockage_x} (left column), together with their corresponding mean profiles $\langle x\rangle_c(t)$ (right column), for three different values of the temperature difference $\Delta T$ (hence different P\'eclet numbers). 
On the left, the solid line represents the frontier for the necessary condition \eqref{eq:cond_block_x}: blockage may arise whenever the colloids cross this line. 
As seen in subfigures \ref{fig:blockage_x}\,$a$, $c$ and $d$, this situation arises for all three values of the temperature difference. 
Indeed, when looking at the corresponding averaged profiles $\langle x\rangle_c(t)$ (subfigures \ref{fig:blockage_x}\,$b$, $d$ and $e$), we observe a small decrease at short times, associated with a short blockage. 
However, only in the third experiment, corresponding to the largest temperature difference, does the blockage last in the long term, showing a mean profile $\langle x\rangle_c(t)$ roughly constant. 

This can be explained by calculating the velocity drift $v_m$ (salt attraction) in the salt-in configuration. 
Using equation \eqref{eq:Vm^i_x_theo} with $i=1$, we obtain:
   \begin{equation}
    v_m(x,t)\sim D_\mathrm{dp} \sqrt{\mathit{Pe}_s} \,\frac{x}{\langle x^2\rangle_s(t)} \underset{t\rightarrow0}{\sim} D_\mathrm{dp} \sqrt{\mathit{Pe}_s} \,\frac{x}{\langle x^2\rangle_s(0)}\, .
   \label{Vm_x_theo_saltin}
    \end{equation}
Therefore, at short time, the salt attraction is more important at higher P\'eclet number. 
On the other hand, the common transport (reference case) is roughly independent of the P\'eclet number, as seen in Section \ref{sec:reference}. 
Since the colloids will all the more get blocked as they remain long enough in their cell, the blockage is more likely to happen when the transient salt effect is strong enough to cancel the common transport and therefore more probable at a larger P\'eclet number. 

\begin{figure}
    \includegraphics{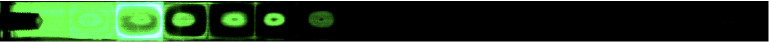}
    \caption{Blockage situation. for $\Delta T=1.45\,\degreecelsius$ and $t=5400\,\second=1.5\,\hour$ (dashed line in Fig. \ref{fig:blockage_x}\,$e$). The coefficient $R$ increases from left to right, following equation \eqref{eq:coeff_blockage_R_in}.
    }
    \label{fig:blockage_image}
\end{figure}

Finally, note that when colloids are blocked, even though the blockage condition \eqref{eq:cond_block_x} is no longer satisfied (figure \ref{fig:R_salt_in}), they remain blocked until diffusion has had time to bring them near the walls again. 
When looking at figure \ref{fig:blockage_image} which shows a snapshot of the experiment when blockage is reached, with the small diffusion coefficient of the colloids, it can clearly take a long time, which explains why the quantity $\langle x\rangle_c(t)$ in figure \ref{fig:blockage_x}$f$ remains roughly constant for so long. 

\section{Summary and conclusion}

We have studied experimentally the joint transport of salt and colloids in a steady Rayleigh-Bénard flow composed of an alignment of steady convection rolls in a closed horizontal chamber. 
By means of High Dynamical Range Light Induced Fluorescence, we have measured the instantaneous concentration fields of colloids and salt to address how they are transported at long times. 
In such configuration, we recover diffusive-like transport at long time for the salt \citep{solomon1988passive,bib:shraiman1987,soward_1987,young1989anomalous} : after a long transient following injection of salt on one side of the chamber, the coarse-grained salt concentration profile reaches a Gaussian shape at large scale, and spreads with a diffusive scaling, with sharp gradients located between convection cells. 
Note that we do not recover diffusive-like dispersion in the case of colloids alone (without salt) due to their interaction with the metallic walls, an effect which was found to be nearly independent on the P\'eclet number.

In order to address the influence of the salt gradients onto colloids transport and test the theoretical predictions derived in \citet{bib:volk_etal2022phoresis}, we have tested two configurations which were compared to the reference case of colloids transport without diffusiophoresis.

In the salt-out configuration, for which colloids and salt are injected on both ends of the chamber with a tunable time delay, the salt gradient is moderate and the blockage criterion adapted from \citet{bib:volk_etal2022phoresis} is rather low. We find that the transport of colloids is enhanced as compared to the reference case, as predicted in the theoretical work. 
Moreover, we obtain a qualitative agreement between the measured spreading rate of the colloids at intermediate times and estimations derived, assuming the coarse-grained salt concentration is Gaussian at large scale and scaling laws derived in \citet{bib:volk_etal2022phoresis} can be transposed to the present situation. 
However, the prediction based on diffusiophoresis falls off in the long time regime since this effect can be only transient in a cell of finite size.

In the salt-in configuration, for which colloids and salt are injected together in the chamber, strong phoretic effects can occur at short time as salt has not entirely spread over the chamber.
For the three increasing values of the colloids P\'eclet number, we observe a transient blockage of the colloids in convection cells, the longer the trapping the larger the P\'eclet number. 
The occurrence of such trapping is found in agreement with the threshold for blockage derived from the salt profile; increasing the P\'eclet number enhances the transient salt attraction, and can annihilate the effect of the second mechanism (colloids attracted by the horizontal walls) which plays against blockage.

Although the results obtained in \citet{bib:volk_etal2022phoresis} were derived with strong hypotheses (constant gradient, stationary regime, linear velocity drift), the theory proved to give satisfactory predictions in the present experimental case for which none of these hypotheses were strictly holding. 
This shows that scale separation, whether in time or space, or the precise nature of the phoretic mechanism, are less important than the topology of the flow (with closed streamlines) to observe the regimes of enhanced transport or blockage. As a consequence the present results and analysis should hold in a large variety of transport phenomena, for instance transport of living cells in cellular flows under the influence of chemotaxis.

\section*{Acknowledgments}
This work was supported by the French research programs ANR-16-CE30-0028. 
For the purpose of Open Access, a CC-BY public copyright licence has been applied by the authors to the present manuscript and will be applied to all subsequent versions up to the Author Accepted Manuscript arising from this submission.

\section*{Declaration of Interests.}
The authors report no conflict of interest.

%\appendix

%\section{Theoretical prediction of colloidal dispersion in the salt-out case}

%\bibliographystyle{unsrt}
\bibliographystyle{jfm}

\bibliography{Compressibility}% Produces the bibliography via BibTeX.

\end{document}